\documentclass[fleqn,reqno,a4paper,12pt]{amsart}

\usepackage[utf8]{inputenc}
\usepackage{amsmath}
\usepackage{amsfonts}
\usepackage{amssymb}
\usepackage[left=2cm,top=3cm,right=2cm,bottom=3cm]{geometry}

\begin{document}

\title[ A   Fifth-Order Bi-Hamiltonian System ]{A   Fifth-Order Bi-Hamiltonian  System  }

\author[Daryoush. Talati]{Daryoush Talati}
 \email{daryoush.talati@eng.ankara.edu.tr,~~daryoush.talati@gmail.com}

\address{ No. 15, 22nd Alley, Jomhoori Eslami Boulevard, Salmas, Iran}

\newtheorem{theo}{Theorem}[section]
\newtheorem{prop}{Proposition}

\begin{abstract}

In this work, we introduced a new two component fifth-order bi-Hamiltonian system admitting the scalar Kupershmidt equation as a reduction.

\end{abstract}

\maketitle

\section*{Introduction}

To prove the integrability of an equation  suspected to be bi-Hamiltonian, one need to find an appropriate compatible pair of Hamilton operator $J$ and $K$ such that the Magri scheme

\begin{equation*}
u_{t_i} =F_i[u]=KG_i[u]=JG_{i+1}[u]  ,~ i=-1,0,1,2,3,...
\end{equation*}

constructed by  the operators contains the equation in hand. Here $F_i[u]$ are characteristics of symmetries and $G_i$ are the conserved gradients.

The scalar equations of order up to 5 are extensively classified with respect to existence of sufficiently many higher conserved densities for the existence of a formal symmetry. Unlik lower order cases, classification of  fifth order two-component evolution equations is an obstinate problem. Some completely integrable equations of this type were found by  Mikhailov, Novikov and Wang \cite{mnw1,mnw2} in the study of symbolic representation theory and  nonevolutionary equations


\begin{equation}
\left(
\begin{array}{l}
u\\ \\
v\\ \\
\end{array}\right)_{t}=
\left(
\begin{array}{cc} 
-\frac{5}{3}u_{5x} -10vv_{3x}+10uu_{3x}+ 25u_{x}u_{xx}  - 15v_{x}v_{xx}  -12u^2u_{x}\\+6v^2u_{x}+12uvv_{x}-6v^2v_{x}\\
15v_{5x} -10vu_{3x}-30uv_{3x}-35v_{x}u_{xx} +30v_{x}v_{xx} -45u_{x}v_{xx}\\ +6v^2u_{x}-6v^2v_{x} + 12uvu_{x} + 12u^2v_{x}  

\end{array}
\right) ,\label{sys1}
\end{equation}

\begin{equation}
\left(
\begin{array}{l}
u\\ 
v\\ 
\end{array}\right)_{t}=
\left(
\begin{array}{cc} 
u_{5x} + 10uu_{3x} + 25u_{x}u_{xx}+20u^2u_{x}+v^2v_{x}\\
u_{3x}v + u_{xx}v_{x}+8uvu_{x}+4u^2v_{x}
\end{array}
\right) ,\label{sys2}
\end{equation}


\begin{equation}
\left(
\begin{array}{l}
u\\ 
v\\ 
\end{array}\right)_{t}=
\left(
\begin{array}{cc} 
-\frac{1}{8}u_{5x}-2 uu_{3x}-2u_{x}u_{xx}-\frac{32}{5}u^2u_{x}+v_{x}\\
\frac{9}{8}v_{5x} + 6uv_{3x} + 6u_{x}v_{xx} + 4u_{xx}v_{x}+ \frac{32}{5}u^2v_{x}
\end{array}
\right) .\label{sys3}
\end{equation}


System (\ref{sys1}) and (\ref{sys2}) admit a reduction $v=0$ to the Kaup-Kupershmidt equation $u_t=u_{5x}+10uu_{3x}+25u_{x}u_{xx}+20u^2u_{x}$ and By setting $v=0,$ system (\ref{sys3}) reduces to the Sawada-Kotera equation   $u_t=u_{5x}+5uu_{3x}+ 5u_{x}u_{xx}+5u^2u_{x}$ (see \cite {mss,w} and references therein).

Bi-Hamiltonian structures   for  (\ref{sys2}) and  (\ref{sys3})  can be found in \cite{mnw2}. Bi-Hamiltonian structure for  system (\ref{sys1}) and Zero curviture representation for (\ref{sys2})  are discussed in \cite{v,s}.
 Very recently system (\ref{sys2}) considered by De Sole, Kac and Turhan \cite{dkt} in the study of the Lenard-Magri scheme of integrability who developed a new method based on the notion of strongly skew-adjoint differential operators, using the Lie superalgebra of variational polyvector fields. However, no two-component  completely integrable  system with  reduction $v=0$ to the kupershmidt equation is known so far. In this work, we introduce a new  system of this type whose bi-Hamiltonian structure we constructed too.

\section*{The new System}

The new fifth order bi-Hamiltonian two-component system we introduce here is

\begin{equation}
\left(
\begin{array}{l}
u\\\\\\\\\\\\\\
v\\\\\\\\\\\\
\end{array}\right)_{t}=
\left(
\begin{array}{cc} 
u_{5x}-30vv_{4x}+5u_{x}u_{3x}-5u^{2}u_{3x}+15v^{2}u_{3x}-75v_{x}v_{3x}+60uvv_{3x}\\+90v^{2}v_{3x}+5u_{xx}^{2}-20uu_{x}u_{xx}+60vv_{x}u_{xx}-45v_{xx}^{2}+90vu_{x}v_{xx}\\+90uv_{x}v_{xx}+540vv_{x}v_{xx}+30u^{2}vv_{xx}-180uv^{2}v_{xx}-90v^{3}v_{xx}\\-5u_{x}^{3}+45u_{x}v_{x}^{2}+60uvu_{x}v_{x}-180v^{2}u_{x}v_{x}+5u^{4}u_{x}\\-90u^{2}v^{2}u_{x} +45v^{4}u_{x}+180v_{x}^{3}+30u^{2}v_{x}^{2}-360uvv_{x}^{2}\\ -270v^{2}v_{x}^{2} -60u^{3}vv_{x}+180uv^{3}v_{x} \\ \\
-9v_{5x}+10vu_{4x}+25v_{x}u_{3x}+20uvu_{3x}+30v^{2}u_{3x}+15u_{x}v_{3x}+90v_{x}v_{3x}\\+15u^{2}v_{3x}+15v^{2}v_{3x}+30u_{xx}v_{xx}+50vu_{x}u_{xx}-10u^{2}vu_{xx}+50uv_{x}u_{xx}\\+60vv_{x}u_{xx}+60uv^{2}u_{xx}+30v^{3}u_{xx}+90v_{xx}^{2}+60uu_{x}v_{xx}+60vv_{x}v_{xx}\\+45u_{x}^{2}v_{x}-20uvu_{x}^{2}+60v^{2}u_{x}^{2}-10u^{2}u_{x}v_{x}+90v^{2}u_{x}v_{x}\\-20u^{3}vu_{x} +120uvu_{x}v_{x}+60uv^{3}u_{x}+15v_{x}^{3}-5u^{4}v_{x}\\+90u^{2}v^{2}v_{x} -45v^{4}v_{x} .

\end{array}
\right) .\label{new}
\end{equation}
By setting $v=0$  the well known Kupershmidt equation is an abvious reduction of system (\ref{new}):
\vspace*{-0.3cm}
\begin{equation*}
u_t=u_{5x}+5u_{x}u_{3x}+5u_{xx}^2-5u^2u_{3x}-20uu_{x}u_{xx}-5u_{x}^3+5u^4u_{x}.
\vspace*{-1cm}
\end{equation*}
 

\begin{prop}
System (\ref{new}) can be written in Hamiltonian form in not just one but two different ways: 
\begin{equation}
\left(
\begin{array}{cc} 
u_{t}\\
v_{t}
\end{array}
\right)
=
F_1 [u,v]
=
\mathrm{J}
\left(
\begin{array}{cc} 
\delta_{u}\\
\delta_{v}
\end{array}
\right)
\int \rho_{1}~ \mathrm{d}x
=\mathrm{K}
\left(
\begin{array}{cc} 
\delta_{u}\\
\delta_{v}
\end{array}
\right)
\int \rho_{-1} ~\mathrm{d}x\label{hm}
\end{equation}

with the campatible pair of Hamiltonian operators

\begin{equation*}
\mathrm{J}= \left(
\begin{array}{cc} 
 3D_x & 0  \\
 0 & D_x 
\end{array}\right),\mathrm{K} = \left(
\begin{array}{cc} 
\mathrm{K} _{1}&\mathrm{K} _{2}\\
-\mathrm{K}^{*} _{2}&\mathrm{K} _{4}
\end{array}\right) 
\end{equation*}

where

\begin{eqnarray*}
\begin{array}{ll}
K_{1}=& 2 D_x^7 +  \alpha_{1} D_x^5 + D_x^5 \alpha_{1} +  \alpha_{2} D_x^3 + D_x^3 \alpha_{2} + \alpha_{3} D_x + D_x \alpha_{3}
+4   u_{x}D_x^{-1}u_{t} +4   u_{t}D_x^{-1}u_{x}\\\\
K_{2} =&-56 D_x^6 v + D_x^5 \alpha_{4} + D_x^4 \alpha_{5} + D_x^3 \alpha_{6} + D_x^2 \alpha_{7} + D_x \alpha_{8} + \alpha_{9}
+4   u_{x}D_x^{-1}v_{t} +4   u_{t}D_x^{-1}v_{x}  \\\\
K_{4}=& -18 D_x^7 +  \alpha_{10} D_x^5 + D_x^5 \alpha_{10} +  \alpha_{11} D_x^3 + D_x^3 \alpha_{11}  + \alpha_{12} D_x + D_x \alpha_{12}
 +4   v_{x}D_x^{-1}v_{t} +4   v_{t}D_x^{-1}v_{x}  \\\\

\end{array}
\end{eqnarray*}
 
where

\begin{eqnarray*}
\begin{array}{ll}

\alpha_{1}=&6(u_{x}-u^{2}+11v^{2})\\\\

\alpha_{2}=&-16u_{3x}+20uu_{xx}-336vv_{xx}+29u_{x}^{2}-6u^{2}u_{x}-381v_{x}^{2}+9u^{4}-294u^{2}v^{2}+264uvv_{x}\\&
+45v^{4}+234v^{2}u_{x}-180v^{2}v_{x}\\\\

\alpha_{3}=&2(5u_{5x}+102vv_{4x}-25u_{x}u_{3x}+3u^{2}u_{3x}-117v^{2}u_{3x}+453v_{x}v_{3x}-72uvv_{3x}+90v^{2}v_{3x}\\&
-21u_{xx}^{2}-600vv_{x}u_{xx}+8uu_{x}u_{xx}-8u^{3}u_{xx}+264uv^{2}u_{xx}+351v_{xx}^{2}-498vu_{x}v_{xx}+540vv_{x}v_{xx}\\&-306uv_{x}v_{xx}+174u^{2}vv_{xx}-180uv^{2}v_{xx}-90v^{3}v_{xx}+6u_{x}^{3}-498u_{x}v_{x}^{2}+180v_{x}^{3}+264u^{2}v_{x}^{2}\\&
-360uvv_{x}^{2}-270v^{2}v_{x}^{2}-2u^{6}+60u^{4}v^{2}-60u^{3}vv_{x}-44u^{2}u_{x}^{2}-90u^{2}v^{4}+360u^{2}v^{2}v_{x}\\&
-6uu_{4x}+180uv^{3}v_{x}+1116uvu_{x}v_{x}+294v^{2}u_{x}^{2})\\\\

\alpha_{4}=&4(53v_{x}+28uv+42v^{2})\\\\

\alpha_{5}=&2(-165v_{xx}-192vu_{x}-212uv_{x}-312vv_{x}+32u^{2}v-168uv^{2}-72v^{3})\\\\

\alpha_{6}=&4(66v_{3x}+141vu_{xx}+165uv_{xx}+90vv_{xx}+281u_{x}v_{x}+45v_{x}^{2}-32u^{3}v-37u^{2}v_{x}-6u^{2}v^{2}\\&
+72uv^{3}-86uvu_{x}+312uvv_{x}-18v^{4}+246v^{2}u_{x}+177v^{2}v_{x})\\\\

\alpha_{7}=&2(-54v_{4x}-194vu_{3x}-264uv_{3x}-36vv_{3x}-557v_{x}u_{xx}+168uvu_{xx}-462v^{2}u_{xx}-603u_{x}v_{xx}\\&
+162v_{x}v_{xx}+57u^{2}v_{xx}-360uvv_{xx}-63v^{2}v_{xx}-288v^{3}u_{x}+184vu_{x}^{2}-1224vu_{x}v_{x}+108uv^{2}u_{x}\\&
+306uu_{x}v_{x}-180uv_{x}^{2}-246vv_{x}^{2}+148u^{3}v_{x}+24u^{3}v^{2}+336u^{2}vu_{x}+24u^{2}vv_{x}-708uv^{2}v_{x}\\&
-36v^{3}v_{x}-4u^{4}v+72uv^{4}+36v^{5})\\\\

\alpha_{8}=&2(9v_{5x}+46vu_{4x}+189v_{x}u_{3x}-80uvu_{3x}+138v^{2}u_{3x}+249u_{x}v_{3x}-90v_{x}v_{3x}-15u^{2}v_{3x}\\&
+72uvv_{3x}-15v^{2}v_{3x}+300u_{xx}v_{xx}-90v_{xx}^{2}-174uu_{x}v_{xx}+360vu_{x}v_{xx}-120vv_{x}v_{xx}-114u^{3}v_{xx}\\&
-324uv_{x}v_{xx}+126uv^{2}v_{xx}-190u_{x}^{2}v_{x}+180u_{x}v_{x}^{2}-30v_{x}^{3}+8u^{5}v+4u^{4}v_{x}+36u^{3}vu_{x}\\&
-48u^{3}vv_{x}-456u^{2}u_{x}v_{x}-192u^{2}v^{2}u_{x}-60u^{2}v^{2}v_{x}-174u^{2}vu_{xx}+108uv_{4x}-204uv_{x}u_{xx}\\&
-72uv^{5}-60uv^{3}u_{x}+72uv^{3}v_{x}-144uv^{2}u_{xx}-308uvu_{x}^{2}-168uvu_{x}v_{x}+492uvv_{x}^{2}-72v^{4}u_{x}\\&
+114v^{3}u_{xx}-84v^{2}u_{x}^{2}+648v^{2}u_{x}v_{x}+180v^{2}v_{x}^{2}-222vu_{x}u_{xx}+564vv_{x}u_{xx})\\\\

\alpha_{9}=&4(-9uv_{5x}+v_{x}u_{4x}+10uvu_{4x}+20u^{2}vu_{3x}+25uv_{x}u_{3x}+30uv^{2}u_{3x}+15u^{3}v_{3x}+15uu_{x}v_{3x}\\&
+90uv_{x}v_{3x}+15uv^{2}v_{3x}-30vv_{x}v_{3x}+5u_{x}v_{x}u_{xx}+30uu_{xx}v_{xx}+90uv_{xx}^{2}-45v_{x}^{2}v_{xx}\\&
+60u^{2}u_{x}v_{xx}+120uvv_{x}v_{xx}+90v^{2}v_{x}v_{xx}+30u^{2}vv_{x}^{2}+30vu_{x}v_{x}^{2}+30uv_{x}^{3}-4u^{5}v_{x}\\&
-20u^{4}vu_{x}-10u^{3}u_{x}v_{x}+60u^{3}v^{2}v_{x}-10u^{3}vu_{xx}+45u^{2}v_{x}u_{xx}+60u^{2}v^{3}u_{x}+60u^{2}v^{2}u_{xx}\\&
-20u^{2}vu_{x}^{2}+120u^{2}vu_{x}v_{x}+40uu_{x}^{2}v_{x}+30uv^{3}u_{xx}+60uv^{2}u_{x}^{2}+90uv^{2}u_{x}v_{x}\\&
-180uv^{2}v_{x}^{2}+50uvu_{x}u_{xx}+60uvv_{x}u_{xx}-90v^{3}v_{x}^{2}+15v^{2}v_{x}u_{xx}+180vv_{x}^{3})\\\\

\alpha_{10} =&2(9u_{x}+54v_{x}+9u^{2}+13v^{2}) \\\\

\alpha_{11} =&-36u_{3x}-288v_{3x}-72uu_{xx}+72vu_{xx}-68vv_{xx}-81u_{x}^{2}-36u_{x}v_{x}-18u^{2}u_{x}+144uvu_{x}\\&
+62v^{2}u_{x}-275v_{x}^{2}-36u^{2}v_{x}-192v^{2}v_{x}-9u^{4}+62u^{2}v^{2}-69v^{4} \\\\

\alpha_{12} =&2(9u_{5x}+90v_{5x}+18uu_{4x}-36vu_{4x}+26vv_{4x}+81u_{x}u_{3x}-54v_{x}u_{3x}+9u^{2}u_{3x}-72uvu_{3x}-51v^{2}u\\&
_{3x}+18u_{x}v_{3x}+239v_{x}v_{3x}+18u^{2}v_{3x}+96v^{2}v_{3x}+63u_{xx}^{2}+54uu_{x}u_{xx}-216vu_{x}u_{xx}-108uv_{x}u_{xx}\\&
-154vv_{x}u_{xx}+18u^{3}u_{xx}-102uv^{2}u_{xx}-60v^{3}u_{xx}+231v_{xx}^{2}-92vu_{x}v_{xx}+396vv_{x}v_{xx}-92u^{2}vv_{xx}\\&
+108v^{3}v_{xx}+18u_{x}^{3}-92v^{2}u_{x}^{2}-108u_{x}^{2}v_{x}+54u^{2}u_{x}^{2}-62u_{x}v_{x}^{2}+192v_{x}^{3}\\&
+20u^{2}v^{2}u_{x}-120uv^{3}u_{x}-308uvu_{x}v_{x}-60v^{4}u_{x}+384v^{2}v_{x}^{2}+18v^{6}+10u^{4}v^{2}\\&
-62u^{2}v_{x}^{2}-60u^{2}v^{4}) 
\end{array}
\end{eqnarray*}

\end{prop}

By straigthforward calculation it is easy to show that  the functional trivector of  linear combination $K+\lambda J$ with constant $\lambda$, vanishes independently from the value of $\lambda$ \cite{OLV}.

The first few conserved densities of the hierarchy are listed below.


\begin{eqnarray*}
\begin{array}{ll}

\rho_{-1}=&\alpha\\
\rho_{0}=&u^2 + 3v^2\\
\rho_{1}=&3uu_{4x}-81vv_{4x}-5u^{2}u_{3x}+45v^{2}u_{3x}-90uvv_{3x}-270v^{2}v_{3x}-5u^{3}u_{xx}+45uv^{2}u_{xx}+180v^{3}u_{xx}\\&
+135u^{2}vv_{xx}+45v^{3}v_{xx}+60u^{3}vv_{x}-540u^{2}v^{2}v_{x}-540uv^{3}v_{x}+u^{6}\\&-45u^{4}v^{2}+135u^{2}v^{4}-27v^{6}\\
\\

\rho_{2}=&-90uu_{6x}+7290vv_{6x}+126u^{2}u_{5x}-756v^{2}u_{5x}+13608uvv_{5x}+20412v^{2}v_{5x}+105u^{3}u_{4x}\\&-5670uv^{2}u_{4x}
-3780v^{3}u_{4x}+11340uv_{x}v_{4x}-11340u^{2}vv_{4x}-34020uv^{2}v_{4x}+16065v^{3}v_{4x}\\&-15120uvv_{x}u_{3x}-70u^{4}u_{3x}
+10080u^{2}v^{2}u_{3x}-7560uv^{3}u_{3x}-2835v^{4}u_{3x}-136080uvv_{x}v_{3x}\\&-7140u^{3}vv_{3x}+45360u^{2}v^{2}v_{3x}
+11340uv^{3}v_{3x}-17010v^{4}v_{3x}+315u^{2}u_{xx}^{2}-11340uvu_{xx}v_{xx}\\&+18900u^{2}vv_{x}u_{xx}-22680uv^{2}v_{x}u_{xx}
-252u^{5}u_{xx}+8820u^{3}v^{2}u_{xx}+13608v^{5}u_{xx}-5670u^{2}v_{xx}^{2}\\&-68040uvv_{xx}^{2}-76545v^{2}v_{xx}^{2}
-6300u^{3}v_{x}v_{xx}+136080u^{2}vv_{x}v_{xx}+34020uv^{2}v_{x}v_{xx}\\&+6930u^{4}vv_{xx}+13608v^{5}v_{xx}+7560u^{3}v^{2}v_{xx}
-22680u^{2}v^{3}v_{xx}-4410u^{4}v_{x}^{2}+15120u^{3}vv_{x}^{2}\\&-34020u^{2}v^{2}v_{x}^{2}+1512u^{5}vv_{x}
-11340u^{4}v^{2}v_{x}-68040u^{2}v^{4}v_{x}-68040uv^{5}v_{x}+30u^{8}\\&-1260u^{6}v^{2}+11340u^{2}v^{6}-2430v^{8}\\\\
\dot{.}&\\
\dot{.}&\\
\end{array}
\end{eqnarray*}

These densities suffice to write two  Magri shemes with same hamiltonian operators that one of them contains the new system proving integrability of the system (\ref{new}).

The recursion relation 

\begin{equation}
F_n
=
\mathrm{J}
\left(
\begin{array}{cc} 
\delta_{u}\\
\delta_{v}
\end{array}
\right)
\int \rho_{n}~ \mathrm{d}x
=\mathrm{K}
\left(
\begin{array}{cc} 
\delta_{u}\\
\delta_{v}
\end{array}
\right)
\int \rho_{n-2} ~\mathrm{d}x
\end{equation}

is a more general relation of (\ref{hm}) that will provide further conservation laws for n-th stage symmetry of the original system.

let us also mention that the recursion operator can be written in the form $R=KJ^{-1}$.  Starting with the basic symmetries $ \left(
\begin{array}{cc} 
u_x\\
v_x
\end{array}
\right)$ and $ \left(
\begin{array}{cc} 
u_t\\
v_t
\end{array}
\right)$ we can generate two infinite generalized symmetries  by applying

\begin{equation}
K_{n+2}=RK_n~~.
\end{equation}

\end{document}